\def\beq{\begin{equation}}
\def\eeq{\end{equation}}

\def\rfr#1{eq. (\ref{#1})}

\def\virg#1{``#1''}

\def\eqi{\begin{equation}}
\def\eqf{\end{equation}}
\def\eqia{\begin{eqnarray}}
\def\eqfa{\end{eqnarray}}
\def\rp#1#2{{#1\over#2}} \def\lb#1{\label{#1}}

\def\virg#1{``#1''}

\documentclass[11pt]{article}
\usepackage{amsmath,amsthm,amscd,amssymb}
\usepackage{latexsym}
\usepackage{graphicx,epsfig}

\begin{document}

\noindent{\bf \LARGE{Constraining the Kehagias-Sfetsos solution of the Ho\v{r}ava-Lifshitz modified gravity with  extrasolar planets}}
\\
\\
\\
{L. Iorio$^{\ast}$, M.L. Ruggiero$^{\S,||}$}\\
{\it $^{\ast}$Ministero dell'Istruzione, dell'Universit\`{a} e della Ricerca (M.I.U.R.). Fellow of the Royal Astronomical Society (F.R.A.S.). Address for correspondence: Viale Unit$\grave{a}$ di Italia 68
70125 Bari (BA), Italy.  \\ e-mail: lorenzo.iorio@libero.it}\\
{\it $^{\S}$ UTIU, Universit\`a Telematica Internazionale Uninettuno, Corso Vittorio Emanuele II 39, 00186 - Roma, Italy. \\
$^{||}$ Dipartimento di Fisica, Politecnico di Torino, Corso Duca degli Abruzzi 24, 10129 - Torino, Italy. \\ e-mail: matteo.ruggiero@polito.it}

\vspace{4mm}

\begin{abstract}
We consider  the  Kehagias-Sfetsos (KS) solution in  the Ho\v{r}ava-Lifshitz  gravity that is the analog of the general relativistic Schwarzschild black hole.  In the weak-field and slow-motion approximation, we, first, work out the correction to the third Kepler law of a test particle induced by such a solution. {Then, we}  compare it to the phenomenologically determined orbital periods of the transiting extrasolar planet HD209458b \virg{Osiris} to preliminarily obtain an order-of-magnitude lower  bound on the KS dimensionless parameter  $\omega_0\geq 1.4\times 10^{-18}$. As suggestions for further analyses, the entire data set of HD209458b should be re-processed by explicitly modeling KS gravity as well, and one or more dedicated solve-for parameter(s) should be estimated.
\end{abstract}


\section{Introduction} \label{sec:intro}
The recently proposed model of quantum gravity  by  Ho\v{r}ava \cite{horava1,horava2,horava3} has recently attracted much attention, and many aspects of it
 have been extensively analyzed, ranging from formal developments, cosmology, dark energy
 and dark matter, spherically symmetric solutions, gravitational waves, and its viability with observational constraints; for a full list of references see, e.g., \cite{cinesi,Bom,Wang,Carloni}.
Such a  theory admits  Lifshitz's  scale invariance: $\mathbf{x} \rightarrow b \mathbf{x}, \quad t \rightarrow b^{q} t$, and, after this, it is referred to as Ho\v{r}ava-Lifshitz (HL) gravity.  Actually, it has anistropic scaling in the short distances domain (UV), since it is $q=3$, while isotropy is recovered at large distances (IR).

One of the key features of the theory is its good UV behavior, since it is power-counting renormalizable; for a discussion of the renormalizability beyond power counting arguments, see  \cite{Orla}. However, in its original formulation, it experiences some problems: for instance, it leads to a non-zero cosmological constant with the wrong sign, in order to be in agreement with the observations \cite{Soda,horatius,visser1}. To circumvent these issues, it was suggested to abandon the principle of ``detailed balance'' \cite{Calca1,Calca2}, initially introduced by Ho\v{r}ava  in his model to restrict the number of possible parameters. As a consequence, phenomenologically viable extensions of the theory were proposed \cite{visser1,visser2}. It was also shown that HL gravity can reproduce General Relativity (GR) at large distances \cite{pope,KS}; for other solutions non-asymptotically flat see \cite{Cai1,Cai2}. However, there is still an ongoing discussion on the consistency of HL gravity, since it seems that modes arise which develop exponential instabilities at short distances, or become strongly coupled \cite{padilla,blas}. Moreover, according to  \cite{Miao}, the constraint algebra does not form a closed structure. Perturbative instabilities affecting HL gravity have been pointed out in  \cite{Saridakis}.

Actually, it is important to stress that, up to now, in HL gravity the gravitational field is purely geometrical: in other words, the way matter has to be embedded still needs to be studied. Nevertheless, there are interesting vacuum solutions that can be studied, such as the static spherically symmetric solution found by Kehagias and Sfetsos (hereafter KS) \cite{KS}. Such a  solution is the analog of Schwarzschild solution of GR and, moreover, it asymptotically reproduces the usual behavior of Schwarzschild spacetime. It is interesting to point out that it is obtained without requiring the projectability condition, assumed in the original HL theory, while spherically symmetric solutions with the projectability condition are however available \cite{tang,Wang}. {Nonetheless, because of its simplicity, it is possible to consider  KS solution as toy model useful to better understand some phenomenological implications of HL gravity.  Actually, in \cite{Tib} it was shown that KS solution  is in agreement with the classical tests of GR, while in a previous paper \cite{HLIR} we studied the corrections to the general relativistic Einstein's pericentre precession determined by this solution and compared the theoretical predictions to
the latest determinations of the corrections  to the standard Newtonian/Einsteinian planetary perihelion precessions recently estimated with the EPM2008  ephemerides.  We found that the KS dimensionless parameter is constrained from the bottom at $\omega_0\geq 10^{-12}-10^{-24}$ level depending on the planet considered.

In our analysis, we assumed that particles followed geodesics of KS metric: however, it is important to point out that this is true if matter is minimally and universally coupled to the metric, which is not necessarily true in HL gravity, where, as we said above, the role of matter has not been yet clarified. In this paper, starting from the same assumption, we focus on the effects induced by the examined solution on the orbital period $P_{\rm b}$ of a test particle, on an extra solar system environment. We will explicitly work out the consequent correction $P_{\omega_0}$ to the usual third Kepler law in Section \ref{3kep}. In Section \ref{osiris} we compare it with the observations of the transiting extrasolar planet HD209458b \virg{Osiris}. {We point out that the resulting constraints are to be considered as preliminary and just order-of-magnitude figures because, actually, the entire data set of HD209458b should be re-processed again by explicitly modeling the effect of the KS gravity; however, this is outside the scopes of the present paper.} Section \ref{conclu} is devoted to the conclusions.
\section{KS corrections to the third Kepler law}\lb{3kep}
As shown in \cite{HLIR},
from \cite{Brum}
\begin{eqnarray} \ddot x^i &=& -\frac{1}{2}c^2 h_{00,i} - \frac{1}{2}c^2 h_{ik}h_{00,k}+h_{00,k}\dot x^k\dot x^i \nonumber \\ &+& \left(h_{ik,m}-\frac{1}{2}h_{km,i}\right)\dot x^k\dot x^m,\ i=1,2,3,\end{eqnarray}
it is possible to obtain the following radial acceleration acting upon a test particle at distance $d$ from a central body of mass $M$
\eqi \vec{A}_{\omega_0}\approx \frac{4 (GM)^4}{\omega_0 c^6 d^5}\hat{d},\lb{accel}\eqf valid up to terms of order $\mathcal{O}(v^2/c^2)$.
Its effect on the pericentre of a test particle have been worked out in  \cite{HLIR}; here we want to look at a different orbital feature affected by  \rfr{accel} which can be compared to certain observational determinations.
%

The mean anomaly is defined as
\eqi \mathcal{M}\doteq n(t-t_p);\eqf in it $n=\sqrt{GM/a^3}$ is the Keplerian mean motion, $a$ is the semimajor axis and $t_p$ is the time of pericentre passage.
The anomalistic period $P_{\rm b}$ is the time elapsed between two consecutive pericentre passages; for an unperturbed Keplerian orbit it is $P_{\rm b}=2\pi/n$. Its modification due to a small perturbation of the Newtonian monopole can be evaluated  with  standard perturbative approaches. The Gauss equation for the variation of the mean anomaly is, in the case of a radial perturbation $A_d$ to the Newtonian monopole \cite{Ber},
\eqi \frac{d{\mathcal{M}}}{dt}=n-\rp{2}{na}A_d\left(\rp{d}{a}\right)+\rp{(1-e^2)}{nae}A_d\cos f,\lb{Gaus}\eqf where $e$ is the eccentricity and $f$ is the true anomaly counted from the pericentre position.
The right-hand-side of \rfr{Gaus} has to be valuated onto the unperturbed Keplerian orbit given by (see \cite{Roy})
\eqi d=\rp{a(1-e^2)}{1+e\cos f}.\eqf
By using (see \cite{Roy})
\eqi df = \left(\rp{a}{d}\right)^2(1-e^2)^{1/2}d\mathcal{M}\eqf
and
\eqi \int_0^{2\pi} (1+e\cos f)^2\left[2-\rp{(1+e\cos f)}{e}\cos f\right]df=\pi\left(1+\rp{5}{4}e^2\right),\eqf it is possible to work out the correction to the Keplerian period due to \rfr{accel}; it is
\eqi P_{\omega_0}=\rp{4\pi (GM)^4\left(1+\rp{5}{4}e^2\right)}{\omega_0 c^6 n^3 a^6(1-e^2)^{5/2}}=\rp{4\pi (GM)^{5/2}\left(1+\rp{5}{4}e^2\right)}{\omega_0 c^6  a^{3/2}(1-e^2)^{5/2}}.\lb{PHL}\eqf
Note that \rfr{PHL} retains its validity in the limit $e\rightarrow 0$ becoming equal to
\eqi P_{\omega_0}\rightarrow\rp{4\pi (GM)^{5/2}}{\omega_0 c^6  d^{3/2}}\lb{circo},\eqf where $d$ represents now the fixed radius of the circular orbit. It turns out that \rfr{circo} is equal to the expression that can be easily obtained by equating the centripetal acceleration $\Omega^2 d$, where $\Omega$ is the particle's angular speed, to the total gravitational acceleration $GM/d^2 - 4(GM)^4/\omega_0c^6 d^5$ with the obvious assumption that the Newtonian monopole is the dominant term in the sum.
\section{Confrontation with the observations}\lb{osiris}
In the scientific literature there is a large number of papers (see, e.g., \cite{Tal,Wri,Ove,Jae,Rey,Ser1,Ser2,Ior,Bro,Peri,Mof,Adl,Page}) in which the authors use the third Kepler law to determine, or, at least,  constrain un-modeled dynamical effects of mundane, i.e. due to the standard Newtonian/Einsteinian laws of gravitation, or non-standard, i.e. induced by putative modified models of gravity.
As explained below, in many cases such a strategy has been, perhaps, followed in a self-contradictory way, so that the resulting  constraints on, e.g., new physics, may be regarded as somewhat \virg{tautologic}.

Let us briefly recall that the orbital period $P_{\rm b}$ of two point-like bodies of mass $m_1$ and $m_2$ is, according to the third Kepler law,
\eqi P^{\rm Kep}=2\pi\sqrt{\rp{a^3}{G{M}}},\eqf
where $a$ is the relative semi-major axis and ${M}=m_1+m_2$ is the total mass of the system.
Let us consider an unmodeled dynamical effect which induces a non-Keplerian (NK) correction to the third Kepler law, i.e.
\eqi P_{\rm b}= P^{\rm Kep} + P^{\rm NK},\eqf
where
\eqi P^{\rm NK}=P^{\rm NK}(M,a,e; p_j),\eqf
is the analytic expression of the correction to the third Kepler law  in which $p_j$, $j=1,2,...$N are the parameters of the NK effect to be determined or constrained. Concerning standard physics, $P_{\rm NK}$ may be due to the centrifugal oblateness of the primary, tidal distortions, General Relativity; however, the most interesting case is that in which $P_{\rm NK}$ is due to some putative modified models of gravity. {As a first, relatively simple step to gain insights into the NK effect one can act as follows.}
By comparing the measured orbital period to the computed Keplerian one it is possible, in principle, to obtain preliminary information on the dynamical effect investigated from $\Delta P\doteq P_{\rm b}-P^{\rm Kep}$. {Actually, one should re-process the entire data set of the system considered by explicitly modeling the non-standard gravity forces, and simultaneously solving for one or more dedicated parameter(s) in a new global solution along with the other ones routinely estimated. Such a procedure would be, in general, very time-consuming and should be repeated for each models considered. Anyway, it is outside the scopes of the present paper, but it could be pursued in further investigations.}

Concerning our simple approach, in order to meaningfully solve for $p_j$ in
\eqi\Delta P=P^{\rm NK}\eqf
it is necessary that
\begin{itemize}
  \item In the system considered a measurable quantity which can be identified with the orbital period  and directly measured independently of the third Kepler law itself, for example from spectroscopic or photometric measurements, must exist. This is no so obvious as it might seem at first sight; indeed, in a N-body system like, e.g., our solar system a directly measurable thing like an \virg{orbital period} simply does not exist  because the orbits of the planets are not closed due to the non-negligible mutual perturbations. Instead, many authors use values of the \virg{orbital periods} of the planets which are retrieved just from the third Kepler law itself.
      Examples of systems in which there is a measured orbital period are many transiting exoplanets, binaries and, e.g,  the double pulsar.
      Moreover, if the system considered follows an eccentric path one should be careful in identifying the measured orbital period with the predicted sidereal or  anomalistic periods. A work whose authors are aware of such issues is \cite{Capoz}.
  \item  The quantities entering $P^{\rm Kep}$, i.e. the relative semimajor axis $a$ and the total mass $M$, must be known independently of the third Kepler law. Instead, in many cases values of the masses obtained by applying just the third Kepler law itself are used.
      Thus, for many exoplanetary systems the mass $m_1\doteq M_\bigstar$ of the hosting star should be taken from stellar evolution models and the associated scatter should be used to evaluate the uncertainty $\delta M_\bigstar$ in it, while for the mass $m_2=m_p$ of the planet a reasonable range of values should be used instead of straightforwardly taking the published value because it comes  from the mass function which is just another form of the third Kepler law.
       {Some extrasolar planetary systems represent good scenarios because it is possible to know many of the parameters entering $P^{\rm Kep}$ independently of the third Kepler law itself, thanks to the redundancy offered by the various techniques used.}

Such issues have been accounted for in several astronomical and astrophysical scenarios in, e.g.,  \cite{IorWD,IorIJMPA,IorCyg,IorReg,IorNA}.
\end{itemize}

\section{The transiting exoplanet HD209458b}
Let us consider HD 209458b \virg{Osiris}, which is the first exoplanet\footnote{See on the WEB http://www.exoplanet.eu/} discovered with the transit method \cite{Cha,Hen}. Its orbital period $P_{\rm b}$ is known with a so high level of accuracy that it was proposed  to use it for the first time to test General Relativity in a planetary system different from ours \cite{osi}; for other proposals to test General Relativity with different orbital parameters of other exoplanets, see  \cite{ada,ung1,ung2,RAGO}.

In the present case, the system's parameters entering the Keplerian period
i.e. the relative semimajor axis $a$, the mass $M_{\bigstar}$ of the host star and the mass $m_p$ of the planet, can be determined independently of the third Kepler law itself, so that it is meaningful to compare the photometrically measured orbital period $P_{\rm b}=3.524746$ d \cite{exo} to the computed Keplerian one $P^{\rm Kep}$: their difference can be used to put genuine constraints on KS solution which predicts the corrections of \rfr{PHL} to the third Kepler law.
Indeed, the mass $M_{\bigstar}=1.119\pm 0.033$M$_{\odot}$ and the radius $R_{\bigstar}=1.155^{+0.014}_{-0.016}$R$_{\odot}$ of the star \cite{exo}, along with other stellar properties, are fairly straightforwardly estimated by matching direct spectral observations with stellar evolution models since for HD 209458 we have also the Hipparcos parallax $\pi_{\rm Hip}=21.24\pm 1.00$ mas \cite{Perry}.
The semimajor axis-to-stellar radius ratio $a/R_{\bigstar}=8.76\pm 0.04$ is estimated from the photometric light curve, so that $a=0.04707^{+0.00046}_{-0.00047}$AU \cite{exo}. The mass $m_p$ of the planet can be retrieved from the parameters of the photometric light curve and of the spectroscopic one entering the formula for the planet's surface gravity $g_p$  (eq.(6) in \cite{exo}). As a result, after having computed the uncertainty in the Keplerian period by summing in quadrature the errors due to $\delta a,\delta M_{\bigstar},\delta m_p$, it turns out
 \eqi \Delta P\doteq P_{\rm b}-P^{\rm Kep} = 204\pm 5354\ {\rm s};\lb{exop}\eqf the uncertainties $\delta M_{\bigstar}$, $\delta a$,  $\delta m_p$  contribute  4484.88 s, 2924.77 s, 2.66 s, respectively to $\delta(\Delta P)=5354$ s.

 The discrepancy $\Delta P$ between $P_{\rm b}$ and $P^{\rm Kep}$ of \rfr{exop} is statistically compatible with zero; thus, \rfr{exop} allows to constrain the parameter $\omega_0$ entering $P_{\omega_0}$.
 Since
 \eqi P^{\rm NK}\doteq P_{\omega_0}=\rp{\mathcal{K}}{\omega_0},\eqf
 with
 \eqi \mathcal{K}\doteq\rp{4\pi(GM)^{5/2}}{c^6 d^{3/2}}=8\times 10^{-15}\ {\rm s},\lb{cazza}\eqf
 by equating the non-Keplerian correction $P_{\omega_0}$ to the measured $\Delta P$ one has
 \eqi \omega_0=\rp{\mathcal{K}}{\Delta P}.\lb{vaffa}\eqf
 Since $\Delta P$ is statistically compatible with zero, the largest value of $\omega_0$ is infinity; from \rfr{vaffa} a lower bound on $|\omega_0|$ can be obtained amounting to\eqi |\omega_0|\geq 1.4\times 10^{-18}.\lb{lowerbo}\eqf
 A confrontation with the solar system constraints\footnote{To avoid confusions with the perihelion $\omega$, the KS parameter is dubbed $\psi_0$ in  \cite{HLIR}.}  Our previous paper \cite{HLIR} shows that such a lower bound is at the level of those from Jupiter and Saturn, while it contradicts the possibility that values of $\omega_0$ as small as those allowed by Uranus, Neptune and Pluto ($|\omega_0|\geq 10^{-24}-10^{-22}$) may exist.
 However, tighter constraints are established by the inner planets for which $|\omega_0|\geq 10^{-15}-10^{-12}$.

\section{Conclusions}\lb{conclu}
We have investigated how the third Kepler law is modified by  the KS solution,
whose Newtonian and lowest order post-Newtonian limits coincides with those of GR, by using the standard Gauss perturbative approach. The resulting expression for $P_{\omega_0}$, obtained from the Gauss equation of the variation of the mean anomaly $\mathcal{M}$, in the limit $e\rightarrow 0$ reduces to the simple formula which can be derived by equating the centripetal acceleration to the Newton$+$KS gravitational acceleration for a circular orbit.

Then, after having discussed certain subtleties connected, in general, with a meaningful use of the third Kepler law to put on the test alternative theories of gravity, we compared our explicit expression for $P_{\omega_0}$ to the discrepancy $\Delta P$ between the phenomenologically determined orbital periods $P_{\rm b}$ and the computed Keplerian ones $P^{\rm Kep}$ for the transiting extrasolar planet HD209458b \virg{Osiris}. Since $\Delta P$ is statistically compatible with zero, it has been possible to {preliminary} obtain the lower bound $|\omega_0|\geq 1.4 \times 10^{-18}$ on the  dimensionless KS parameter. {However, the entire data set of HD209458b should be re-processed by including KS gravity as well, and a dedicated, solve-for parameter should be estimated as well. The previously reported } constraint rules out certain smaller values allowed by the lower bounds obtained from the perihelia of Uranus, Neptune and Pluto ($|\omega_0|\geq 10^{-24}-10^{-22}$). On the other hand, our exoplanet bound still leaves room for values of $\omega_0$ too small according to the constraints from the perihelia of Mercury, Venus and the Earth ($|\omega_0|\geq 10^{-14}-10^{-12}$).


\end{document}